# Ultrafast electron energy-loss spectroscopy in transmission electron microscopy

Enrico Pomarico, Ye-Jin Kim, F. Javier García de Abajo, Oh-Hoon Kwon, Fabrizio Carbone, and Renske M. van der Veen

Enrico Pomarico, École Polytechnique Fédérale de Lausanne, Switzerland; enrico.pomarico@epfl.ch

Ye-Jin Kim, Ulsan National Institute of Science and Technology, South Korea; gne1019@unist.ac.kr

F. Javier García de Abajo, Institut de Ciencies Fotoniques, Spain; javier.garciadeabajo@nanophotonics.es

Oh-Hoon Kwon, Department of Chemistry, Ulsan National Institute of Science and Technology, South Korea; ohkwon@unist.ac.kr

Fabrizio Carbone, École Polytechnique Fédérale de Lausanne, Switzerland; fabrizio.carbone@epfl.ch

Renske M. van der Veen, Department of Chemistry, and Frederick Seitz Materials Research Laboratory, University of Illinois at Urbana-Champaign, USA; renske@illinois.edu

In the quest for dynamic multimodal probing of a material's structure and functionality, it is critical to be able to quantify the chemical state on the atomic-/nanoscale using element-specific electronic and structurally sensitive tools such as electron energy-loss spectroscopy (EELS). Ultrafast EELS, with combined energy, time, and spatial resolution in a transmission electron microscope, has recently enabled transformative studies of photoexcited nanostructure evolution and mapping of evanescent electromagnetic fields. This article aims to describe state-of-the-art experimental techniques in this emerging field and its major uses and future applications.

**Keywords:** electron energy-loss spectroscopy (EELS), laser-induced reaction, nanoscale, electronic structure, optical properties

**Introduction**

Electron energy-loss spectroscopy (EELS) in the transmission electron microscope has become an invaluable tool for unraveling the chemical compositions and structures of materials, enabling imaging of individual atoms and their bonding states with unprecedented resolutions.[1,2] The low-energy ("low-loss," 0–50 eV) region of the EEL spectrum yields electronic information in the form of valence intraband and interband transitions, as well as plasmon excitations, rendering this part of the spectrum sensitive to changes in the overall electron density of the material. Conversely, the high-energy ("core-loss" >100 eV) region of the EEL spectrum is characterized by excitations of core-level electrons into well-defined higher-lying empty states and into the continuum, providing a technique suitable for studying the chemical state, local geometric structure, and nature of chemical bonding centered around the absorbing atom. When combined with the excellent spatial resolution of the transmission electron microscope, EELS constitutes a powerful technique for the electronic characterization of nanoscale materials.

However, if one wishes to study dynamical processes, the temporal resolution has been typically limited by the acquisition time of the detector (~30 ms). Only recently, ultrafast EELS with temporal resolutions ranging from femtoseconds (fs) to nanoseconds (ns) has been demonstrated and applied to study laser-induced pre-ablation dynamics[3] and bandgap renormalization[4] in graphite thin films, electron–phonon coupling and structural dynamics in multiwalled carbon nanotubes,[5] imaging of evanescent waves surrounding carbon nanotubes,[6] plasmonic nanostructures,[7–9] biological structures,[10,11] and photo-induced charge-transfer and phase transitions in transition-metal oxides.[12,13]

This article discusses some of the latest results, prospects for applications, and new methods in ultrafast EELS at the nanometer scale. Refer to References 14–17 for a detailed discussion of ultrafast and dynamic electron microscopy and References 18–20 for a deeper theoretical understanding of inelastic electron–photon interactions.

**Ultrafast core-level electron spectroscopy**

Apart from relativistic effects,[21] electron energy-loss and x-ray absorption core-level spectra are essentially equivalent and provide analogous information. Energy losses between 100 and 1500 eV are routinely accessible in TEM-EELS (TEM, transmission electron microscopy), which (partly) overlaps with the soft x-ray region. Ultrafast x-ray and extreme ultraviolet spectroscopies have experienced tremendous progress in recent years.[22–24] Picosecond-resolved x-ray absorption spectroscopy implemented at synchrotron facilities has enabled the characterization of the excited state structure and dynamics of a wide variety of chemical systems. More recently, the advent of x-ray free-electron lasers has introduced a paradigm shift in terms of the temporal resolution of x-ray spectroscopies into the femtosecond regime. Importantly, due to the rather low interaction cross-section of x-rays with matter, *in situ* studies of nanoscale objects such as thin films and individual nanoparticles are challenging, especially for materials containing low-$Z$ ($Z$ is atomic number) elements such as organic crystals, polymers, and biological molecules. This can be overcome by ultrafast EELS, which combines, in a tabletop apparatus, high spatial resolution and sensitivity to characterize individual nanostructures with the ultrashort temporal resolution and energy resolutions needed to address chemical dynamics immediately following laser excitation.

The feasibility of femtosecond- and nanosecond-resolved core-level EELS has recently been demonstrated by the study of photo-induced structural dynamics in graphite thin films[4] and charge-transfer dynamics in iron oxide photocatalysts.[13] These experiments revealed that core-level EELS is especially sensitive to chemical bonding properties (e.g., structure, symmetry, spin, and charge) in the close vicinity of the absorbing atom. **Figure 1** shows experimental and simulated static and transient energy-loss near-edge spectra (ELNES) at the carbon $K$-edge (~280 eV) after laser excitation at 532 nm. The transient core-loss spectra, in combination with *ab initio* molecular dynamics simulations, reveal the elongation of the carbon–carbon bonds, even though the overall behavior is a thermal contraction of the crystal lattice. A prompt energy-gap shrinkage is observed on the picosecond time scale (Figure 1c), which is caused by local bond length elongation and the direct renormalization of band energies due to temperature-dependent electron–phonon interactions. In femtosecond-resolved core-level EELS, caution is warranted in interpreting the dynamics between $t = -1$ ps and $+1$ ps, where the inelastic photon–electron (photon-induced near-field electron microscopy [PINEM], *vide infra*) effect significantly broadens the spectrum, as shown in Figure 1d. While in some cases careful deconvolution methods are needed to eliminate these features,[13] other experiments may exploit this technique to reveal material-dependent dynamics of plasmonic fields (e.g., due to photo-induced metallicity or electron–hole-pair dynamics).[25]

One of the challenges of ultrafast core-level EELS is that the cross section for inelastic scattering decays with a power-law energy dependence according to $AE^{-r}$, with $r = 2$–$5$, $A$ = constant, and $E$ = energy.[26] Therefore, deep core-level edges (>100 eV) are considerably more challenging to measure than shallow core levels[12] and low-loss plasmon excitations (0–100 eV).[3] In order to overcome these challenges and enhance the sensitivity necessary to measure such weak time-resolved signals at deep core-level ionization edges, careful data acquisition strategies (e.g., energy-drift correction), efficient detection schemes (e.g., direct electron detection[27]), and new instrument developments[28–30] are necessary and already underway.

Compared to ultrafast optical transient spectroscopy, which mainly probes the delocalized/hybridized valence orbitals, or ultrafast electron diffraction, which tracks the overall structural changes of the material, the element specificity and simultaneous local electronic and structural sensitivity of ultrafast core-level EELS opens the door for a plethora of studies on the *L*-edges of transition-metal oxides that are of current interest for microelectronics, spintronics, photocatalytic water splitting, biosensing, and solar-energy conversion. Of particular importance are future studies of hole dynamics, which have remained largely unrevealed using conventional optical-probing techniques, as well as dependences of the

electron–hole dissociation, trapping, and recombination time scales on the nanoparticle size, shape, and defect structure in prevalent photocatalysts.

**Ultrafast low-loss electron spectroscopy with meV resolution**

To characterize and manipulate many-body excitations in the low-energy spectral range of strongly correlated solids, one needs to simultaneously combine meV-spectral, nanometer-spatial, and femtosecond-temporal resolutions. Energy-filtered transmission electron microscopy of phonons and plasmons in nanostructures yields sub-nanometer[31] and 10 s meV space/energy resolution[32–34] under static conditions. To achieve a combined nm-fs-meV resolution, a laser-assisted ultrafast TEM method, where energy resolution is not determined by the electron beam and spectrometer energy spreading, but solely by the laser pulse linewidth, has been proposed.[35] In particular, spectral profiles of plasmonic resonances (PRs) with 20 meV resolution have been retrieved by performing PINEM[6,8,20,36] with infrared (IR) tunable laser excitation combined with quantitative analysis of the obtained energy-filtered images.

When electrons traverse the proximity of a nanostructure, the transient electric field associated with the electrons excites PRs, manifesting themselves as features in an energy-loss spectrum (**Figure 2**a). In ultrafast TEM, to reduce space-charge effects, the intensity of the electron pulses is strongly attenuated. Under these conditions, the energy bandwidth of the pulses is broadened to the eV level, limiting the investigation of features at smaller energy scales. By illuminating the nanostructure with an optical pulse at a resonant energy, a specific PR can be excited and multiples of the PR energy can be exchanged with the imaging electron beam[6,20] (Figure 2b). By scanning the laser wavelength across the PRs and mapping their spatial profile after filtering all inelastically scattered electrons (Figure 2c), one can resolve the modes with an energy resolution only limited by the laser linewidth, and not by the electron-beam energy bandwidth. The spectrum of PRs in an 8 μm-long Ag nanowire has been measured over the energy range between 800 and 1080 meV by using a modified microscope operated at 200 kV[37] in combination with an optical parametric amplifier in the near-IR working at 300 kHz repetition rate.[35] In Figure 2d, we show the measured spectra for *n* = 13–16 modes, which are in rather good agreement with simulations based upon boundary-element-method calculations of the optical field.[38,39]

This method is a valuable tool for providing full dynamical characterization and control of low energy excitations, such as IR-active phonons, without relying on the energy broadening of the incident electron beam. The central frequency and the resolution used to image a specific mode are solely determined by the light-excitation properties. Optical sources of high monochromaticity and controlled temporal profile are readily available at energies covering a wide spectral range and can provide a broader parameter space compared with state-of-the-art electron optics. If one partially sacrifices time resolution, ultrahigh-resolution experiments (even below 1 meV) can be conceived by relying on the intrinsic linewidth of laser sources or spectral shaping techniques.

New fundamental aspects of coupling between different collective excitations, such as phonons and plasmonic fields in a nanostructure, could emerge via this laser-assisted method. For instance, by exciting the IR active phonons in carbon nanotubes[40] optically or by impulsive stimulated Raman scattering, one could investigate the effects on the spatial profile and the spectral changes of the PRs in real time (Figure 2e). Perspectives in chemistry and sensing include the possibility of tracking reaction trajectories and tailoring the chemical sensitivity of nanosensors using this technique.[41] For example, the optical enhancement produced by coupling to plasmons[42,43] could be spatially and spectrally probed in a complete way (Figure 2f). In addition, new opportunities are offered for exploring exotic properties of quantum solids at low energies, such as longitudinal Josephson plasmons (JPs) in layered superconductors,[44] arising from tunneling of Cooper pairs between the superconducting planes. By combining tunable optical excitation with cryo-Lorentz microscopy,[45] one could visualize the dynamical changes of the vortex lattice induced by the external magnetic field upon excitation of a JP (Figure 2g).

**Single-shot EELS to probe irreversible light-induced phenomena**

The stroboscopic laser-pump electron-probe method achieves ~100 fs and sub-eV time and energy resolutions, respectively, by integrating electron counts from many pump-probe cycles in order to achieve sufficient signal in the EEL spectrum. In this scheme, repetitive measurements are made on the same region of a specimen such that observable processes are limited to only highly repeatable, reversible ones, for which the photoexcited specimen must fully return to its original state before successive pump pulses arrive. If a sample undergoes an irreversible process, which is more common in, for example, phase transitions or chemical reactions, the approach should be based on a single laser excitation "shot" and a sequence of subsequent electron pulses that captures the course of the unique irreversible process.[17] The single-shot approach requires short electron probe pulses that are intense enough to generate a spectrum with sufficient signal-to-noise ratio. Dense electron pulses, however, are subject to significant Coulomb repulsion, which not only limits the formation and propagation of the pulses, but also disperses the energy of electrons; pulsed electrons under the space-charge and Boersch effects tend to broaden in space, time, and energy.[46]

First, to satisfy the two extreme conditions of maximized electron counts and minimized energy spread, one needs to optimize the initial state of the photoemitted electron pulse. This can be done by changing the physical shape of the cathode, gun geometry, extraction field, pulse duration, irradiation region, and laser fluence. The replacement of a sharp tip with a flat cathode with large emission area can increase the number of electrons by a few orders of magnitude while preserving similar space-charge effects, but compromising spatial coherence.[47] The control of the gap between the cathode surface and the exit aperture of a Wehnelt cup and the electric field around the aperture can be used to find a configuration for which the number of electrons and their energy spread are optimized.[48–50] The temporal elongation of the pulsed electrons up to a microsecond regime can further increase the number of electrons in a single pulse in the space-charge quasi-free regime by another order of magnitude, reaching a total current of milliamperes, which is equivalent to the steady-state current of thermionic electron sources.

The second technical challenge comprises the dispersion and deflection of a sequence of electron pulses onto different regions of the camera in order to make a "movie" of the ensuing events after laser excitation. Bostangjolo et al.[51] pioneered this "movie-mode" approach, and it was further developed by researchers at Lawrence Livermore National Laboratory.[17,52] This approach is based on inserting a fast, electrostatic deflector into the transmission electron microscope imaging system after the specimen. An electronic timing system switches the voltage of the deflector in the dead time between the incident electron pulses that are generated by a powerful high repetition rate laser system. This technique has been successfully applied to capture images and diffractograms of irreversible processes, such as ultrafast melting[53] and reaction dynamics in nanolaminates,[17,54] with temporal resolutions far exceeding the video-rate millisecond resolution. Its future extension for single-shot EELS studies will merely involve the adaptation of the deflector system to project EEL spectra along the camera direction perpendicular to the dispersion direction of the imaging system. Since EEL spectra are typically only projected onto a fraction of the camera, no compromise in energy resolution is required (in contrast to movie-mode imaging or diffraction that needs to sacrifice resolution).

The methodological concept of single-shot EELS is schematically presented in **Figure 3**. The foreseen characteristics comprise exposure times as short as a few microseconds, time series of up to a dozen pulses with microsecond relative temporal spacing, and energy spreads of a few eV. An increase in energy resolution may be obtained by deconvolution of the intrinsic energy spread using the "zero-loss" peak.[55] While the capabilities of single-shot EELS may appear inferior in terms of temporal and energy resolution compared to the stroboscopic time-resolved EELS techniques, its much wider applicability for studying prevalent irreversible processes, such as photo- and thermally induced phase transformations (including melting) and reactions in chemistry and materials science, encourages its emergence in the near future.[56–58]

**Conclusions**

Ultrafast transmission electron microscopy has emerged as a powerful tool for understanding the evolution of atomic and electronic rearrangements under nonequilibrium conditions in organic, inorganic, and biological materials. Imaging and diffraction alone are incapable of capturing fine modulations in chemical composition and bonding, charge-transfer dynamics, or spin-state changes that often accompany structural reorganizations in materials triggered by external stimuli. Ultrafast EELS therefore complements the crucial fourth dimension (energy) in the dynamic multidimensional visualization of real materials, and transforms a TEM into a true "chemiscope," capable of capturing element-specific snapshots of evolving (nano)structures (e.g., under light, pressure, chemical, or electrical bias), tracking and controlling collective excitations in strongly correlated materials, or mapping plasmonic fields at nanostructure interfaces. This article presented the state of the art and future perspective of a few particularly promising directions in ultrafast EELS, namely ultrafast core-level EELS for the study of transition-metal charge-carrier and spin-state dynamics in heterogeneous catalysis, ultrafast low-loss EELS with meV energy resolution for phonon and plasmon mapping, and single-shot EELS for studies of irreversible phenomena such as phase transitions. Future technological advances in electron sources, acquisition, and detection development are anticipated to enable studies with increased energy resolution and better sensitivity at higher energy losses, as well as new probing schemes for dynamic multimodal probing of the structure and functionality of a material.

**Acknowledgments**

E.P. and F.C. acknowledge support from the SNSF (P300P2_158473) and from the NCCR MUST. O.H.K. and Y.J.K. acknowledge support from the Institute for Basic Science (IBS-R020-D1) and NRF Korea funded by the Ministry of Science, ICT and Future Planning (MSIP) (2017R1A2B4010271). R.M.V. is grateful for startup funds provided by the Department of Chemistry at UIUC.

**Figure captions**

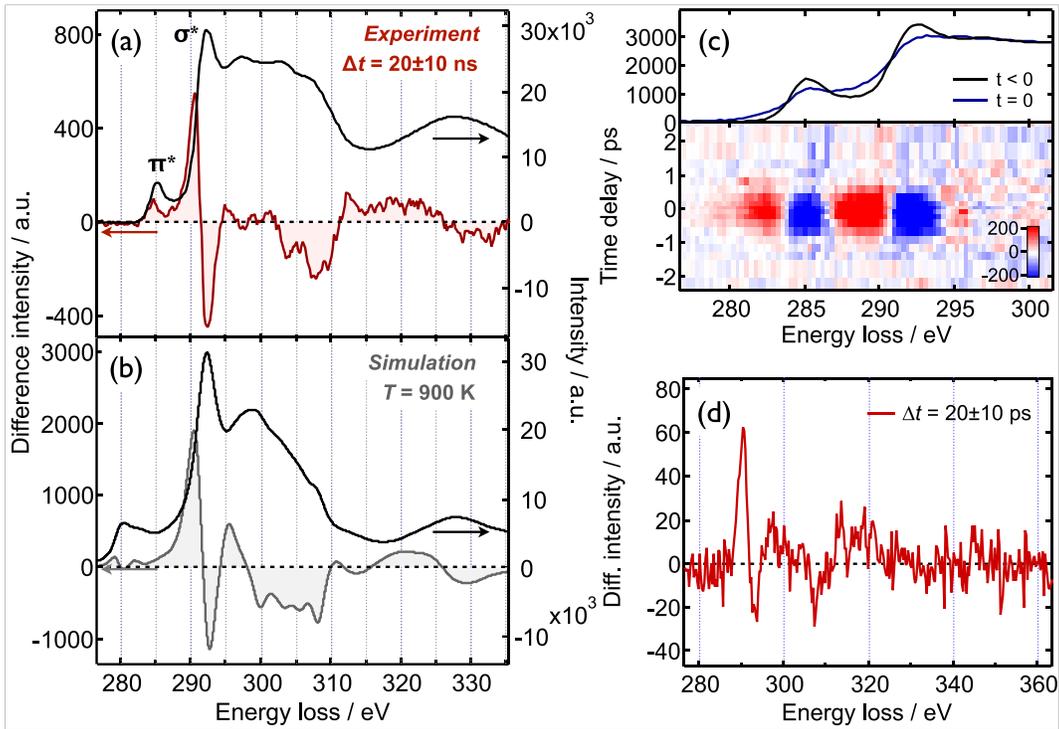

**Figure 1.** (a) Nanosecond-resolved core-level electron energy near-edge spectrum (ELNES) at the carbon *K*-edge before photoexcitation (<$t_0$, right axis) and transient difference spectrum after laser excitation ($\Delta t$ = 10 ns, left axis). (b) Theoretical static and transient ELNES based on molecular dynamics simulations at room temperature and 900 K, respectively. (c) The photon-induced near-field electron microscopy effect in core-level electron energ-loss spectroscopy, in which electron–photon interaction causes a broadening of the spectrum around $t$ = 0. (d) Transient ELNES spectrum at $t = 20 \pm 10$ ps after laser excitation.[4] Note: *t*, time; $t_0$.

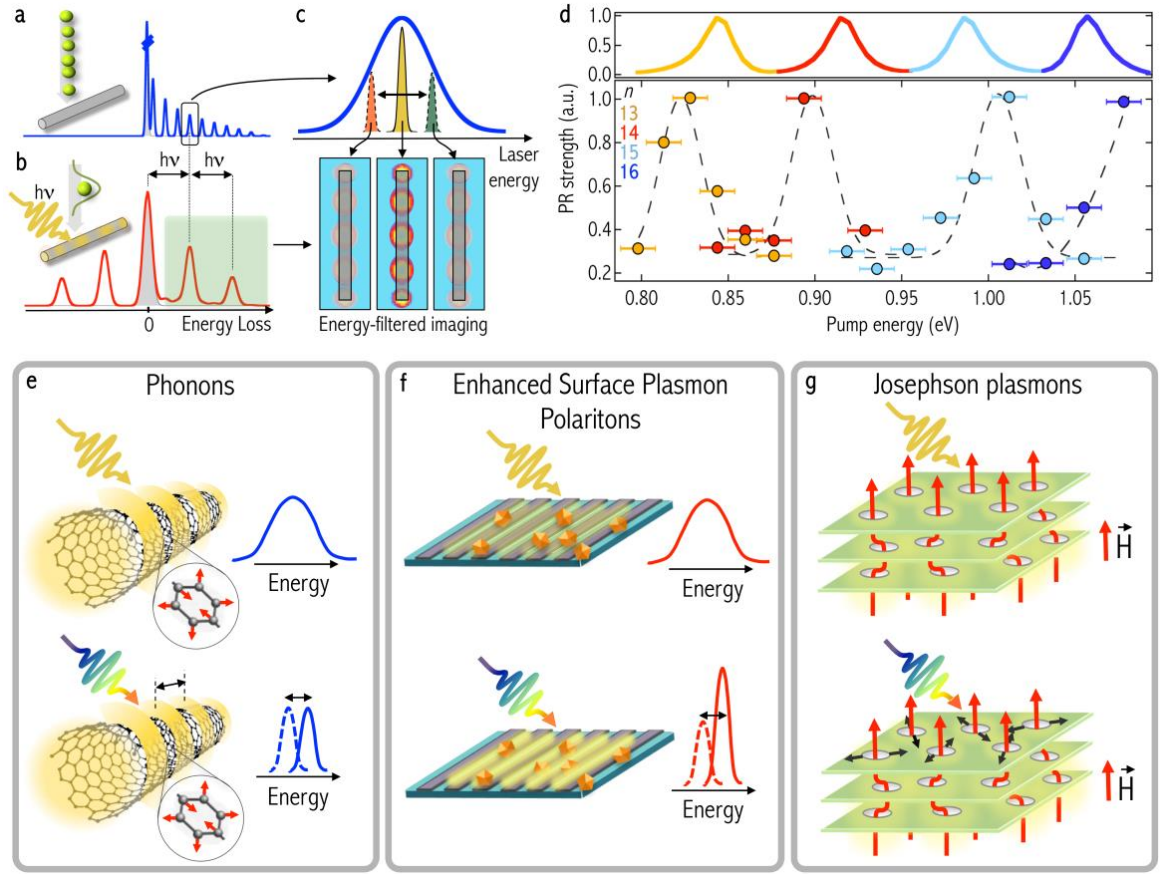

**Figure 2.** (a) Electron energy-loss spectroscopy (EELS) spectrum of plasmonic resonances (PRs) excited in a nanowire by passing electrons. (b) EELS spectrum upon photoexcitation by light pulses of energy $h\nu$ of a specific plasmonic mode, which exchanges its characteristic energy several times with the electrons. (c) The laser excitation wavelength is scanned and the PR profile retrieved via a quantitative analysis of the energy-filtered images. (d) PR strength for modes $n$ = 13–16 obtained by integrating the Fourier transforms of the parallel-to-the-wire spatial profiles of the PRs at specific periodicities and by scanning the laser energy between 0.8 and 1.08 eV (simulations on top). (e–g) Some phenomena that are inaccessible with standard ultrafast energy-filtered transmission electron microscopy (top in each panel), but approachable with the laser-assisted method (bottom in each panel): (e) real-time spatial and spectral changes of the PR features of a carbon nanotube upon excitation of its infrared active phonons; (f) enhancement and spectral shift induced by nanostructures on the surface plasmon polaritons excited on a metallic grating; (g) spatial changes in real-time of the vortex lattice in a layered superconductor upon excitation of a Josephson plasma mode.[35] Note: $h$, Planck's constant; $\nu$, frequency.

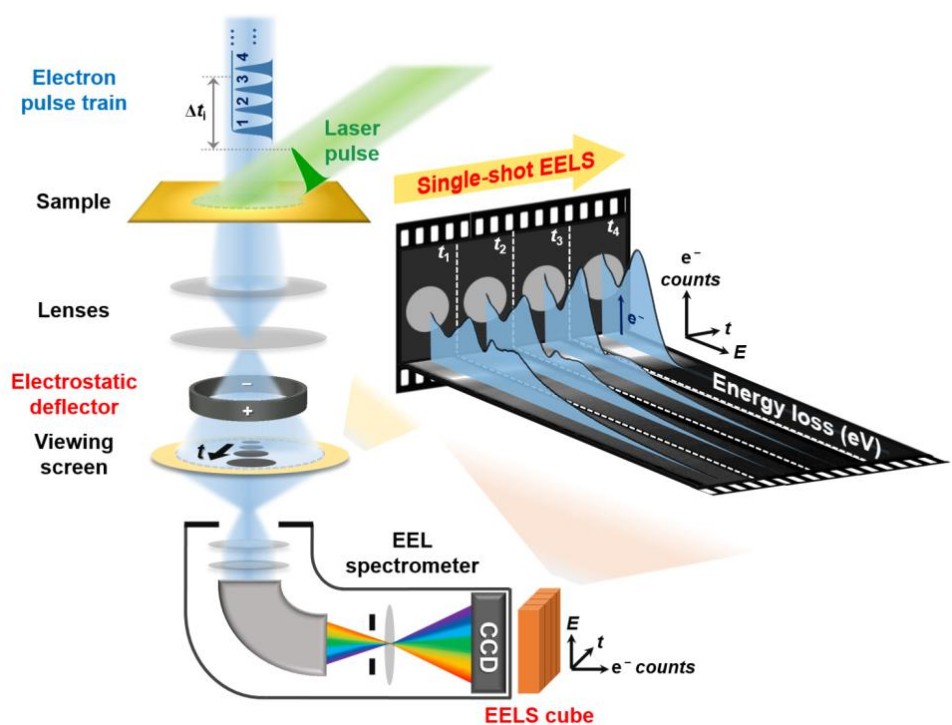

**Figure 3.** The principle of single-shot electron energy-loss spectroscopy (EELS). Structural and electronic dynamics in the specimen are initiated by a single laser pump pulse, after which an electron-probe pulse train captures the light- (or temperature-jump)-induced processes. Each electron-probe pulse is time delayed [$\Delta t$] with respect to the pump pulse, with $\Delta t$ being variable from nanoseconds to microseconds. A fast, electrostatic deflector in the post-specimen column is used to deflect the sequence of electron pulses along the direction perpendicular to the direction of energy dispersion in the EEL spectrometer. An "EELS cube" (energy $E$, time, electron counts) is recorded in a single acquisition enabling the study of irreversible processes. Note: CCD, charge-coupled device.

**Enrico Pomarico** is a research scientist at the École Polytechnique Fédérale de Lausanne (EPFL), Switzerland. He received his PhD degree in quantum optics from the University of Geneva in 2011. He joined the Laboratory of Ultrafast Spectroscopy at EPFL and the Max Planck Institute for the Structure and Dynamics of Matter in Hamburg in 2015. He has studied spin-orbit and electron–phonon coupling effects at short time scales in molecular and solid systems. His current research focuses on combining ultrafast THz optics with transmission electron microscopy to investigate low-energy excitations in quantum solids. Pomarico can be reached by email at enrico.pomarico@epfl.ch.

**Ye-Jin Kim** is a PhD candidate at Ulsan National Institute of Science and Technology, South Korea. She earned her MS degree in 2018 exploring chemical reaction dynamics using time-resolved spectroscopic methods. Her current research focuses on pushing the spatiotemporal resolution of ultrafast electron microscopy to the limit and its application to the study of the ultrafast structural and electronic dynamics of matter at nanoscales. Kim can be reached by email at gne1019@unist.ac.kr.

**F. Javier García de Abajo** is ICREA Professor and Group Leader at the Institut de Ciencies Fotoniques, Spain. He graduated from the University of the Basque Country, completed postdoctoral research at Berkeley National Laboratory, and was a research professor at the Spanish Research Council. He is Fellow of both the American Physical Society and the Optical Society of America. His current interests include surface science, electron microscopy and spectroscopy, nanophotonics, and plasmonics. García de Abajo can be reached by email at javier.garciadeabajo@nanophotonics.es.

**Oh-Hoon Kwon** has been assistant professor in the Department of Chemistry at Ulsan National Institute of Science and Technology (UNIST), South Korea, since 2013. He received his BS degree in 1998 and his PhD degree in 2004, studying picosecond-resolved spectroscopy, at Seoul National University. During postdoctoral training and later as a senior scientist at the California Institute of Technology, he focused on developing the second-generation ultrafast electron microscope (UEM). At UNIST, he is building a high-spatiotemporal resolution UEM for the expansion of its scope to study dynamic phenomena of matter. Kwon can be reached by email at ohkwon@unist.ac.kr.

**Fabrizio Carbone** is a professor at the École Polytechnique Fédérale de Lausanne, Switzerland. He obtained his PhD degree in condensed-matter physics from the University of Geneva in 2007. He then spent two years at the California Institute of Technology conducting ultrafast electron diffraction experiments on organic thin films. His current research focuses on ultrafast optical and electron-based techniques for the investigation of strongly correlated solids and nanostructures. Carbone can be reached by email at fabrizio.carbone@epfl.ch.

**Renske van der Veen** joined the faculty in the Department of Chemistry and the Frederick Seitz Materials Research Laboratory of the University of Illinois at Urbana-Champaign, as an assistant professor in 2015. She obtained her PhD degree from the École Polytechnique Fédérale de Lausanne in 2010 in the field of ultrafast x-ray spectroscopy. Subsequently, she became a postdoctoral researcher at the California Institute of Technology, working on ultrafast electron microscopy and spectroscopy. She was a project group leader at the Max Planck Institute for Biophysical Chemistry in Germany from 2013 to 2015. Her current research focuses on unraveling solar-energy conversion pathways using ultrafast x-ray and electron-based tools. Van der Veen can be reached by email at renske@illinois.edu.